\documentclass[final,journal]{IEEEtran}
\usepackage{amsmath}
\usepackage{amsfonts}
\usepackage{bm}
\usepackage{algorithmic}

\usepackage{algorithm}
\usepackage{multicol}
\usepackage{amssymb}

\usepackage{array}
\usepackage[caption=false,font=normalsize,labelfont=sf,textfont=sf]{subfig}
\usepackage{textcomp}
\usepackage{stfloats}
\usepackage{url}
\usepackage{verbatim}
\usepackage{graphicx}
\usepackage{cite}
\usepackage{color}
\usepackage{cases}
\usepackage{mathtools}
\usepackage{lipsum}
\usepackage{cuted}

\begin{document}
\bstctlcite{IEEEexample:BSTcontrol} %to remove dashes in references with the same authors	
	
	\title{Localization of a Passive Source with a Sensor Network based Experimental Molecular Communication Platform}
	
	\author{
		Fatih Gulec,~\IEEEmembership{Member,~IEEE}, Damla Yagmur Koda, Baris Atakan, Andrew W. Eckford,~\IEEEmembership{Senior Member,~IEEE}
		% <-this % stops a space
		\thanks{FG is with the Department of Electrical Engineering and Computer Science, York University, Toronto M3J 1P3, Canada and was with the Department of Electrical and Electronics Engineering, Izmir Institute of Technology, Izmir, Turkey when most of this study was carried out (Email: fgulec@yorku.ca). DYK is with Vestel Electronics, Manisa, Turkey (Email: damlayagmurkoda@gmail.com). BA is with the Department of Electrical and Electronics Engineering, Izmir Institute of Technology, Izmir, Turkey (Email: barisatakan@iyte.edu.tr). AWE is with the Department of Electrical Engineering and Computer Science, York University, Toronto M3J 1P3, Canada (Email: aeckford@yorku.ca).}
		
		\thanks{The work of FG was supported in part by TUBITAK under Grant 119E041 and in part by a Discovery grant from the NSERC. The work of DYK and BA were supported by TUBITAK under Grant 119E041. The work of AWE was supported by a Discovery grant from the NSERC.}% <-this % stops a space		
	}
	
	
	
\maketitle

\begin{abstract}
	In a practical molecular communication scenario such as monitoring  air pollutants released from an unknown source, it is essential to estimate the location of the molecular transmitter (TX). This paper presents a novel Sensor Network-based Localization Algorithm (SNCLA) for passive transmission by using a novel experimental platform which mainly comprises a clustered sensor network (SN) with $24$ sensor nodes and evaporating ethanol molecules as the passive TX. In SNCLA, a Gaussian plume model is employed to derive the location estimator. The parameters such as transmitted mass, wind velocity, detection time, and actual concentration are calculated or estimated from the measured signals via the SN to be employed as the input for the location estimator. The numerical results show that the performance of SNCLA is better for stronger winds in the medium. Our findings show that evaporated molecules do not propagate homogeneously through the SN due to the presence of the wind. In addition, our statistical analysis based on the measured experimental data shows that the sensed signals by the SN have a log-normal distribution, while the additive noise follows a Student's t-distribution in contrast to the Gaussian assumption in the literature. 
\end{abstract}

\begin{IEEEkeywords}
	Molecular communication,  sensor networks, localization, experimental setup, passive sources.
\end{IEEEkeywords}
	
\section{Introduction}
Molecular communication (MC), which employs chemical signals for information transfer, is a recent and emerging area for communication engineering \cite{atakan2016molecular,nakano2013molecular,farsad2016comprehensive}. The  motivation of MC is that electromagnetic wave-based communication is not suitable for some environments such as human body in microscale \cite{atakan2012body} or wave-denied infrastructures in macroscale \cite{qiu2014molecular, guo2015molecular}. In addition, MC is utilized to model natural phenomena such as infectious disease spread \cite{gulec2021molecular, schurwanz2021duality, gulec2022mobile}, bacterial communication \cite{michelusi2016queuing, gulec2023stochastic, gulec2023stochasticdisruption}, and intercellular plant communication \cite{awan2021modeling}. The applicability of MC systems in such environments makes MC an important communication paradigm for future applications.

In the literature, experimental platforms have been proposed for macroscale MC. The initial experimental platform, serving as a pioneering prototype, included a transmitter (TX) responsible for emitting alcohol molecules through a fan-assisted mechanism. Accompanying this, a receiver (RX) was incorporated into the setup, designed as an alcohol sensor to detect and measure the emitted molecules \cite{farsad2013tabletop}. The data rate performance of the MC system in \cite{farsad2013tabletop} was increased by employing multiple input multiple output  technique \cite{koo2016molecular}. Similarly, a sprayer-based MC system with a camera-based receiver was proposed in \cite{damrath2021investigation}. In \cite{farsad2017novel} and \cite{khaloopour2019experimental}, experimental platforms which encode the information symbols by the pH level of the emitted chemicals were proposed. In \cite{unterweger2018experimental} and \cite{fichera2020fluorescent},  similar systems to the one given in \cite{farsad2017novel} were proposed by using magnetic and carbon nanoparticles instead of chemicals, respectively.  Another experimental platform was accomplished by using an odor generator as the TX and a mass spectrometer as the RX \cite{giannoukos2017molecular,mcguiness2019experimental}. In \cite{abbaszadeh2019mutual} and \cite{atthanayake2018experimental}, flow-based setups were proposed by using particle image velocimetry and planar laser induced fluorescence techniques, respectively. In addition, there are also several experimental MC setups for microscale MC. In \cite{nakano2008microplatform}, a microplatform for intercellular communication by using flourescence imaging was proposed. In \cite{grebenstein2019molecular}, a testbed based on proton pumping bacteria which encode the information by changing the pH level was proposed. Microfluidic platforms were also proposed for bacterial MC \cite{krishnaswamy2013time, amerizadeh2021bacterial}, droplet-based \cite{hamidovic2019information}, and salinity-based MC systems \cite{angerbauer2023}. A graphene-based MC receiver was proposed for a microfluidic platform in \cite{kuscu2021fabrication}. Moreover, a platform consisting of a chemical vapor transmitter and photoionization detectors was proposed in \cite{ozmen2018high}.

The studies employing these experimental platforms are mostly about channel modeling \cite{farsad2014channel, lee2017machine, kim2019experimentally, gulec2021droplet}, efficient data transfer \cite{bhattacharjee2022digital} and distance estimation \cite{gulec2020distance, gulec2021fluid}. However, the localization of a molecular source is important for a more efficient communication. It is also essential to predict the location of a molecular threat such as a virus source propagating through breath in the air \cite{khalid2019communication, xuan2022detection}, or a toxic source which emits evaporating toxic molecules. As for the localization studies in MC literature, methods are proposed mostly for microscale scenarios \cite{etemadi2022abnormality}.  In \cite{kumar2020nanomachine} and \cite{bao2021relative}, maximum likelihood estimation-based methods are proposed to localize the transmitter nanomachine (NM) with two receiver NMs. In \cite{yetimoglu2022multiple}, a method for the localization of two transmitter NMs by using the spatial distribution of received molecules on the surface of the receiver NM is proposed. A triangulation-based method and an iterative gradient descent-based method are proposed to estimate the location of a transmitter NM with multiple receiver NMs reporting to a fusion center in \cite{baidoo2020channel}. In macroscale, theoretical localization methods based on the sensors' detection times processed by the fusion centers absorbing the mobile sensors in a cylindrical fluidic medium are proposed in \cite{khaloopour2021theoretical}. Sensors are assumed to charge linearly their storages when they detect their targets so that the detection time is estimated. In \cite{qiu2015long}, a localization algorithm is proposed by using a mobile search robot as the RX moving towards the source according to molecule concentration gradient  for a long range underwater scenario. However, the performance of these localization methods are not known for practical scenarios. 

All the experimental MC platforms mentioned above primarily revolve around the active transmission of molecules through devices like sprayers or pumps. However, no platform has been developed to understand the dynamics of macroscale MC with passive transmission such as evaporating toxic molecules from a threatening source through the air. Moreover, there is no experimentally validated localization method for practical macroscale scenarios. 

Within this context, we propose a novel experimental platform for macroscale MC applications and a novel localization algorithm using this platform. Firstly, our experimental platform comprises a passive source which includes freely evaporating ethanol molecules. The experimental platform is placed in a fume hood, which is a closed box, to provide controlled conditions. Evaporating molecules are detected by a sensor network (SN) which includes $24$ MQ-3 alcohol sensor nodes in a rectangular arrangement. The novelty of our experimental platform lies in the usage of the SN which can pave the way to novel methods by adapting techniques from the SN literature. Moreover, the concept of employing an SN can be applied for different practical scenarios such as the localization of an underwater molecular TX.

Secondly, the Sensor Network-based Clustered Localization Algorithm (SNCLA)  is proposed for the localization of a passive molecular TX as an application employing our experimental platform. The SN is divided into four clusters. Primarily, the Gaussian plume model which is employed widely in the meteorology literature to model the movement of the pollutant particles in the air is adopted as the system model. As for the SNCLA, the location estimator is derived for the sensor node pairs in each cluster. In order to use the location estimator, some experimental parameters such as the actual concentration, which shows the molecule concentration in kg/m$^3$ instead of the measured  sensor voltage, transmitted mass, and the wind velocity are required to be estimated or calculated. To this end, a detection is made according to the predetermined energy threshold. The measured sensor voltages at the chosen detection threshold are converted to actual concentration values via the sensitivity response of the sensors. The detection times of the SN are employed to estimate the velocity of the wind in the medium for four directions on the $ x-y $ plane. The estimated wind velocity is taken as the input for the mass calculation of the evaporated molecules. The location estimator employs all these estimated/calculated values as the input. Finally, SNCLA determines two clusters according to the magnitude of the wind velocities estimated for the four directions and makes the location estimation for the sensor nodes in these clusters. Numerical results demonstrate that SNCLA yields improved performance under higher wind velocities. Furthermore, the average detection times for all of the sensor nodes are given to show the propagation pattern of the evaporating molecules. It is evident that the dispersal of evaporating molecules is not uniform in all directions, as it becomes apparent that prevailing winds in the medium influence the molecule propagation predominantly in a specific direction aligned with the wind. Moreover, the findings indicate that enhanced precision is achievable by adopting higher detection thresholds.

In comparison to the conference version in \cite{gulec2020localization}, energy detection scheme is employed in SNCLA to improve the localization  performance. Furthermore, a statistical analysis is added by using the experimental data. Thus, it is revealed that the additive noise in the measured sensor signals follows a Student's t-distribution, while the sensed signals have a log-normal distribution. With this perspective the contributions of this paper can be summarized as follows:

\begin{itemize}
	\item An experimental MC platform employing an SN and a passive molecular source is proposed.
	\item The SNCLA is proposed to estimate the location of a passive molecular source with the usage of the proposed experimental platform.
	\item The sensed and noise signals measured by the SN are analyzed and statistically characterized. 
\end{itemize}

The remainder of the paper is organized as follows. In Section \ref{Exp_setup}, the experimental platform is given in detail. Section \ref{System_model} introduces the system model on which the SNCLA is based. The SNCLA  is presented and numerical results are shown in Section \ref{SNCLA}. The statistical analysis is presented in Section \ref{detection}. Finally, the concluding remarks are given in Section \ref{Conclusion}.

\section{Experimental Platform}
\label{Exp_setup}
\begin{figure}[!b]
	\centering
	\includegraphics[width=0.95\columnwidth]{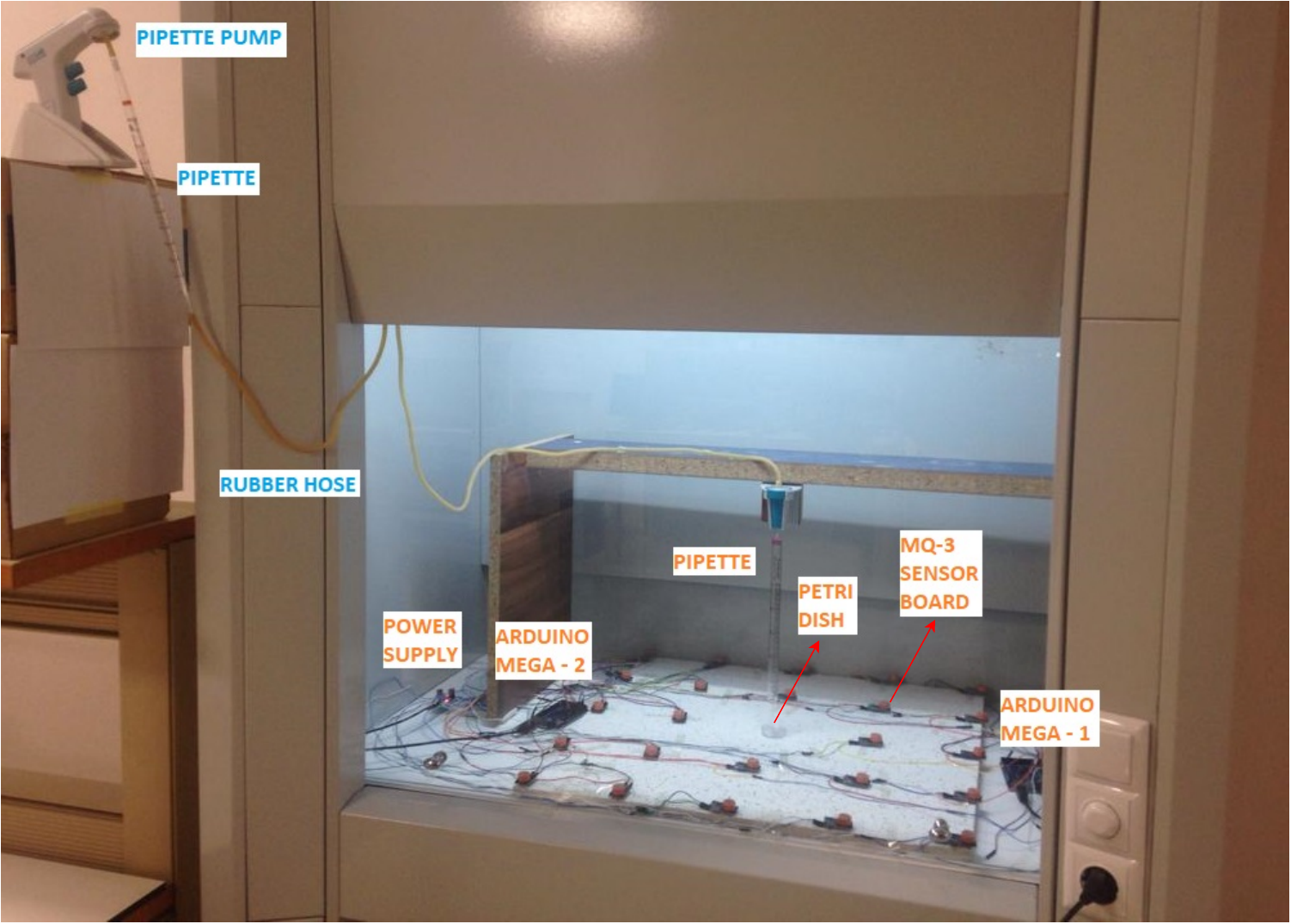}
	\caption{Experimental platform.}
	\label{Setup}
\end{figure}

In this section, the experimental platform which is employed for the localization of a molecular transmitter using an SN is introduced. As shown in Fig. \ref{Setup}, this platform consists of a TX and  an SN placed inside a fume hood, which is a closed cabinet to conduct chemical experiments at controlled conditions without being exposed to chemicals. The TX includes a pipette pump, two pipettes, a rubber hose and a circular petri dish. The pipette connected to the tip of the pipette pump is filled with liquid ethanol before the transmission. When liquid ethanol is pumped through the rubber hose, it fills the petri dish which has a radius of $2.25$ cm. The petri dish is deployed at the midpoint of the SN. The transmission is realized by the evaporation of ethanol molecules in the petri dish at room temperature ($25^\circ C$). After the transmission, evaporated ethanol molecules propagate in the air. 

The SN consists of 24 MQ-3 alcohol sensor boards (or nodes), a power supply and two Arduino Mega microcontroller boards which are connected to a computer. Due to the allocation of $16$ analog input ports within each microcontroller board, the implementation necessitates the integration of two Arduino Mega boards to effectively manage and process the signal inputs originating from the array of $24$ sensor nodes. As shown in Fig. \ref{Setup}, while fourteen of the sensor nodes are wired to the first Arduino microcontroller board, ten of them are connected to the second Arduino board. The sensor nodes are placed on a rectangular surface of $ 60 \times 60$ cm. This rectangular design enabled an easier deployment and wiring of the nodes due to the limited space within the fume hood. The distance between two adjacent nodes on the horizontal and vertical axis is $15$ cm. Each sensor board has a $1$ k$ \Omega$ load resistor on it in order to generate a voltage to be an analog input signal for the microcontroller board.  In order to synchronize the TX and SN, the nodes start to receive signals as soon as the petri dish is filled with $5$ ml of liquid ethanol. Next, the system model to explain the propagation of evaporated molecules employed in the proposed experimental platform is given.
\setcounter{equation}{3}
\begin{figure*}[b]
	\begin{equation}
		C(x,y,z,t) = \frac{ (\pi t)^{-3/2} m_T}{8  (K_x K_y K_z)^{1/2}} \times \text{e}^{\left(-\frac{(x - x_T - u_x t)^2}{4K_xt} - \frac{(y - y_T - u_y t)^2}{4K_yt} \right)}\times \left[ \text{e}^{\left(-\frac{(z - z_T)^2}{4K_zt}\right)} + \text{e}^{\left( - \frac{(z + z_T)^2}{4K_zt} \right)} \right].		\label{atm_disp_eu}
	\end{equation}
\end{figure*}

\begin{figure*}[b]
	\begin{equation}
		C(x,y,z,t) = \frac{m_T}{ (2\pi)^{3/2} \sigma_x \sigma_y \sigma_z} \times\text{e}^{\left(-\frac{(x - x_T - u_x t)^2}{2\sigma_x^2} - \frac{(y - y_T - u_y t)^2}{2\sigma_y^2} \right)} \times \left[ \text{e}^{\left(-\frac{(z - z_T)^2}{2\sigma_z^2}\right)} + \text{e}^{\left( - \frac{(z + z_T)^2}{2\sigma_z^2} \right)} \right]. 
		\label{atm_disp_lang}
	\end{equation}
\end{figure*}
\setcounter{equation}{0}

\section{System Model}
\label{System_model}
This section details the system model on which the localization algorithm is based. In macroscale MC, diffusion-based models are employed to explain the propagation of molecules through the air \cite{farsad2014channel,mcguiness2018parameter,zhai2018anti}. Unlike macroscale experimental studies in the literature, molecules are released by evaporation at room temperature in our scenario. Since there is not any applied force for the emission, we can classify it as a passive transmission. The absence of the force applied to this emission makes molecules susceptible to the effects of wind or flows in the air, even at low velocities. Actually, there is almost always a slight wind in the air \cite{hanna1982handbook}. In the meteorology literature, Gaussian plume models and computational fluid dynamics (CFD) models such as Reynolds averaged Navier Stokes equations coupled with turbulence models are widely employed to model for the dispersion of air pollutants \cite{joseph2020reconciling}. Although CFD models have a higher resolution, their computational cost is higher and they are not analytically tractable. Considering the passive transmission of molecules and winds in the air, Gaussian plume model can be applied for our scenario to model the propagation of evaporated molecules.  By using the conservation of mass, we can write \cite{stockie2011mathematics}

\begin{equation}
	\frac{\partial C }{\partial t} + \nabla \cdot \vec{J} = S,
	\label{cons_mass}
\end{equation} 
where $S$ is the source term, $ \vec{J} $ and $C$ represent the mass flux and concentration of evaporated molecules, respectively. Here, the mass flux can be given as the summation of diffusive flux $ \vec{J_D} $, which stems from the turbulent diffusivity in the atmosphere, and the advective flux $ \vec{J_a} $ stemming from the wind velocity ($ \vec{u} $). Hence, the mass flux is given as
\begin{equation}
	\vec{J}  =  \vec{J_D}  +  \vec{J_a}  = - \vec{K}\nabla C + C\vec{u},
\end{equation}
where $  \vec{K} = \mathrm{diag}(K_x, K_y,K_z)$ is a diagonal matrix showing the turbulent diffusivities in three dimensions. Thus, (\ref{cons_mass}) takes the form of the equation which is known as the atmospheric diffusion (or dispersion) equation as given by \cite{stockie2011mathematics}
\begin{equation}
	\frac{\partial C }{\partial t} + \nabla \cdot (C\vec{u}) = \nabla \cdot (\vec{K}\nabla C) + S.
	\label{atm_eq}
\end{equation}

In our scenario, we define the TX at the position ($x_T, y_T, z_T$) as an instantaneous source  to have a time-dependent solution in (\ref{atm_eq}). In fact, our experimental platform is in a sufficiently small scale so that the TX can be considered as a source releasing molecules in an instantaneous puff.  Given the specific characteristics of our experimental setup, where the dimensions are significantly smaller compared to macroscopic atmospheric scenarios, we adopt this simplification. In this context, the term "sufficiently small scale" refers to the constrained spatial dimensions of our controlled environment, where the evaporation process occurs over a short duration relative to the typical time scales of atmospheric dispersion. Furthermore, the wind velocity is defined with two components in $x$ and $y$ axes, i.e., $u_x$ and $u_y$. It is assumed that the plane at $z=0$ is a reflective plane and there is not any other boundaries. For the source term which is defined as $S = \frac{m_T}{\vec{u}}\delta(x- x_T)\delta(y- y_T)\delta(z- z_T)\delta(t)$ where $m_T$ is the transmitted mass and $\delta(.)$ is the Dirac delta function, the solution of (\ref{atm_eq}) is given in (\ref{atm_disp_eu}) which is known as the Gaussian puff solution \cite{de2013air}. Here, $e^{ - \frac{(z + z_T)^2}{4K_zt} }$ represents the reflection of the plume from the ground. In the literature of atmospheric dispersion, the turbulent diffusivities are defined in terms of dispersion parameters such that $\sigma_x^2 = 2K_x t$, $\sigma_y^2 = 2K_y t$, $\sigma_z^2 = 2K_z t$ \cite{seinfeld2016atmospheric}. Hence, (\ref{atm_disp_eu}) is rearranged as given in (\ref{atm_disp_lang}).

By considering the TX as an instantaneous source, we aim to capture the initial distribution of evaporated molecules efficiently within the experimental time frame. While this approximation deviates from the gradual nature of real-world evaporation processes, it aligns with the practical constraints and time scales of our experiments.

This modeling choice allows us to effectively apply the Gaussian plume model to describe the propagation of evaporated molecules, while acknowledging the simplifications introduced by the constrained scale of our setup. The subsequent sections provide further insights into our dispersion model, its parameters, and the implications of our experimental scale on the modeling assumptions.

The advantage of this conversion  is to determine the dispersion parameters ($\sigma_x$, $\sigma_y$, $\sigma_z$) by using empirically derived models which depend on the distance between the TX and RX. According to the model given in \cite{briggs1973diffusion} which is widely used in the meteorology literature, $\sigma_y$ and $\sigma_z$ for stable air conditions as in our case are calculated by
\setcounter{equation}{5}
\begin{align}
	\sigma_y &= \frac{0.04 r}{(1+0.0001 r)^{0.5}} \label{sigma_y} \\
	\sigma_z &= \frac{0.016 r}{(1+0.0003 r)} 
	\label{sigma_z}
\end{align}
where $r$ is the distance to the source in meters and $\sigma_x$ can be approximated as $\sigma_x \approx \sigma_y$ \cite{de2013air}. Regarding these empirical models in (\ref{sigma_y}) and (\ref{sigma_z}), the effect of the dispersion parameters on the system model are negligible, since the scale of our SN is small ($60\times60$ cm). Therefore, the dispersion parameters are defined as constant values. In our scenario, the SN and TX are all deployed at $z=0$. Accordingly for each sensor, the concentration is given as
\begin{equation}
	\hspace{-0.3cm}C_{i,j}\hspace{-0.1cm} =\hspace{-0.1cm} \frac{m_T \times\text{e}^{\left(\hspace{-0.1cm}-\frac{(x_{i,j}\hspace{-0.1mm} -\hspace{-0.1mm} x_T\hspace{-0.1mm} -\hspace{-0.1mm} u_x t_{i,j})^2}{2\sigma_x^2}\hspace{-0.1mm} -\hspace{-0.1mm} \frac{(y_{i,j}\hspace{-0.1mm} -\hspace{-0.1mm} y_T\hspace{-0.1mm} -\hspace{-0.1mm} u_y t_{i,j})^2}{2\sigma_y^2}\hspace{-0.1cm} \right)}}{\hspace{-0.1mm} \sqrt{2 \pi^3}\hspace{-0.1mm} \sigma_x\hspace{-0.1mm} \sigma_y\hspace{-0.1mm} \sigma_z},
	\label{atm_disp_lang0}
\end{equation}
where $(x_{i,j},y_{i,j})$ and $t_{i,j}$ show the location and detection time of the node $N_{i,j}$ which is in the $i^{th}$ row and $j^{th}$ column of the SN, respectively. Here, $i = {1,...,M_r}$ and $j = {1,...,M_c}$.

\begin{figure}[h]
	\centering
	\includegraphics[width=2.5in]{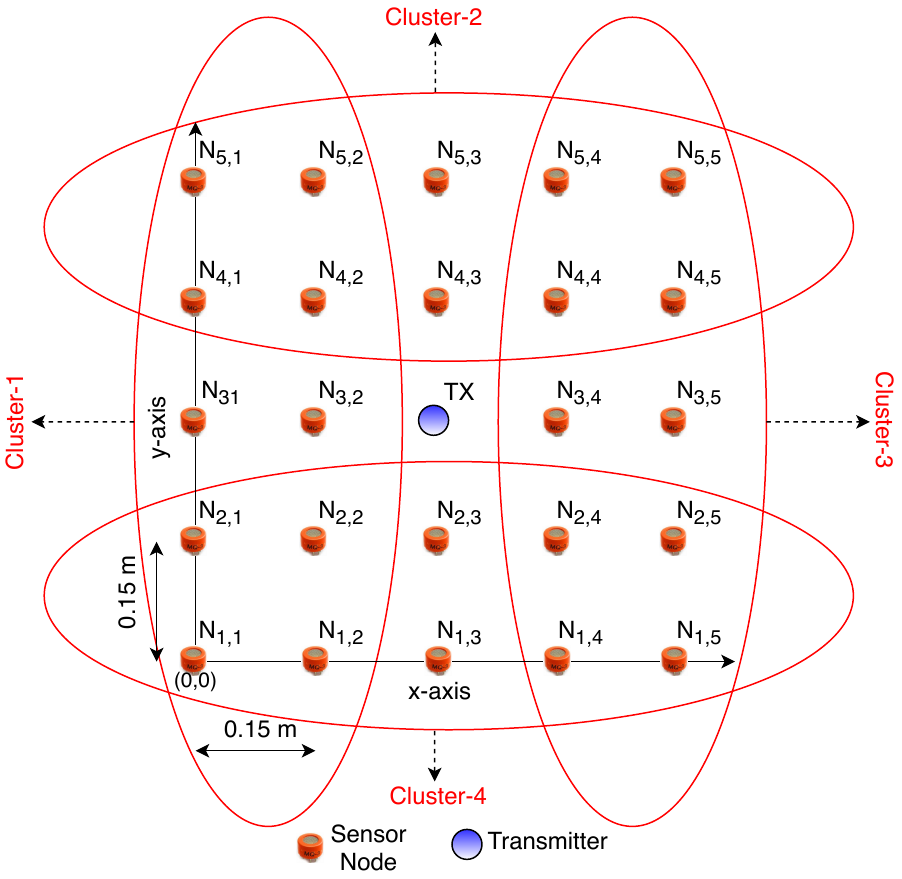}
	\caption{The deployment of the sensor nodes and TX.}
	\label{SN_diagram} \vspace{-0.4cm}
\end{figure}

In addition, the deployment of the sensor nodes are illustrated in Fig. \ref{SN_diagram}. It is assumed that each sensor node knows its location in the Cartesian coordinate system. As shown in this figure, the SN is divided into four clusters. These clusters are employed for the localization algorithm of the TX in MC as given in the next section.

\section{Sensor Network-Based Clustered Localization Algorithm}
\label{SNCLA}
\begin{figure}[h]
	\centering
	\includegraphics[width=\columnwidth]{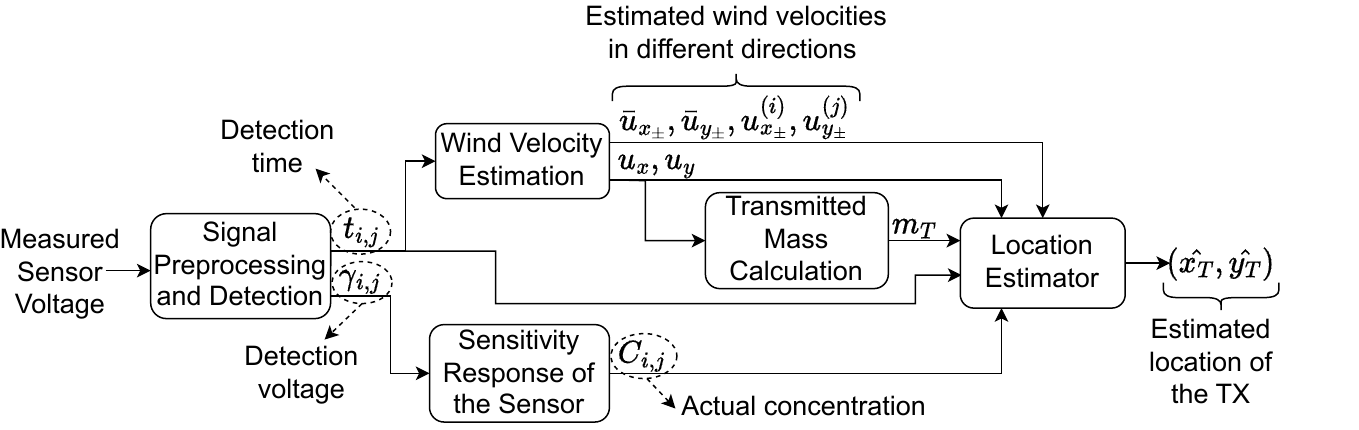}
	\caption{Block diagram of the SNCLA.}
	\label{SN_blocks}
\end{figure}

In this section, Sensor Network-Based Clustered Localization Algorithm \linebreak (SNCLA) whose block diagram is given in Fig. \ref{SN_blocks} is proposed. This block diagram is explained briefly as follows. Initially, the voltage values from the SN are measured and processed to determine the detection time and voltage values via energy detection scheme as the output of the signal preprocessing and detection block. The detection voltages are converted to actual concentration values (kg/m$^3$) in the sensitivity response of the sensor block by using the sensitivity characteristics of the sensor. In addition, detection times are employed to estimate the wind velocities in different directions in the wind velocity estimation block. The estimated wind velocities are then used to calculate the evaporated mass from the source. Finally, the transmitted mass, detection times, actual concentration, and wind velocity values are employed as the input to the location estimator block in which the location of the source is calculated.

Based on the block diagram given in Fig. \ref{SN_blocks}, this section is organized as follows. First, the location estimator is derived using the system model given in Section \ref{System_model}. Then, the estimation and calculation of the required parameters for the location estimator are detailed. At the end of this section, the operation of the SNCLA is detailed by employing all the estimated and calculated parameters. \vspace{-0.4cm}
\vspace{-1cm}
\subsection{Derivation of the Location Estimator}
The motivation to derive the location estimator is to employ the system model which is based on the Gaussian plume model explained in Section \ref{System_model}. In this model, a relation between the measured concentration values and the location of the TX, i.e., $x_T$ and $y_T$, can be established as shown in (\ref{atm_disp_lang0}). Here, our aim is to manipulate (\ref{atm_disp_lang0}) in order to derive an analytical expression and hence to understand which parameters are needed for the location estimation. To this end, (\ref{atm_disp_lang0}) can be written as
\begin{multline}
	\frac{ \sqrt{2\pi^3} \sigma_x \sigma_y \sigma_z C_{i,j}}{m_T} =\\ \text{exp}\left(-\frac{(x_{i,j}\hspace{-0.1cm} -\hspace{-0.1cm} x_T\hspace{-0.1cm} -\hspace{-0.1cm} u_x t_{i,j})^2}{2\sigma_x^2}\hspace{-0.1cm} -\hspace{-0.1cm} \frac{(y_{i,j}\hspace{-0.1cm} -\hspace{-0.1cm} y_T\hspace{-0.1cm}-\hspace{-0.1cm} u_y t_{i,j})^2}{2\sigma_y^2} \right)\hspace{-0.1cm}.
	\label{est_1}
\end{multline}
When the natural logarithm, i.e., $\ln(\cdot)$, of both sides  is taken, then (\ref{est_1}) is given by	
\begin{multline}
	\text{ln}\left( \frac{ \sqrt{2\pi^3} \sigma_x \sigma_y \sigma_z C_{i,j}}{m_T} \right)=-\frac{(x_{i,j} - x_T - u_x t_{i,j})^2}{2\sigma_x^2} \\ - \frac{(y_{i,j} - y_T - u_y t_{i,j})^2}{2\sigma_y^2}.
	\label{est_2}
\end{multline}

For convenience, let $n_{i,j} = \text{ln}\left( \frac{ \sqrt{2}(\pi)^{3/2} \sigma_x \sigma_y \sigma_z C_{i,j}}{m_T} \right)$. Hence, the final equation for the location estimator is given by
\begin{equation}
	\frac{(x_{i,j}\hspace{-0.1cm} -\hspace{-0.1cm} x_T \hspace{-0.1cm}-\hspace{-0.1cm} u_x t_{i,j})^2}{2\sigma_x^2}\hspace{-0.1cm} +\hspace{-0.1cm} \frac{(y_{i,j}\hspace{-0.1cm} -\hspace{-0.1cm} y_T\hspace{-0.1cm} -\hspace{-0.1cm} u_y t_{i,j})^2}{2\sigma_y^2}\hspace{-0.1cm} +\hspace{-0.1cm} n_{i,j} = 0.
	\label{est_3}
\end{equation}

For two sensor nodes, a system of nonlinear equations can be generated using (\ref{est_3}) where $x_T$ and $y_T$ are the variables and the other parameters are constant. Since the solution of this system is not easily tractable, numerical methods can be used to obtain the solution as detailed later in this section. In order to solve these equations, parameters such as $C_{i,j}$, $t_{i,j}$, $m_T$ and wind velocity values are required to be estimated or calculated.
\subsection{Signal Preprocessing and Detection} \label{SPD}
In our experimental platform, the concentration is measured as a voltage value from the sensor nodes. Due to turbulent diffusion of the molecules and electronic noise stemming from the circuit elements in the experimental platform, there are fluctuations on the measured sensor voltage ($V_{out}$). In our previous work \cite{gulec2020localization}, an amplitude detection scheme which includes a moving average filter was used in the first block of the SNCLA given in Fig. \ref{SN_blocks}. In this paper, we employ an energy detection scheme instead of the amplitude detection scheme to increase the accuracy of the localization and reduce the complexity by removing the filter. Both detection schemes are detailed as follows. 

\subsubsection{Amplitude Detection Scheme} In order to detect the signals more accurately, the effect of noise and diffusion of molecules in the received signal by the sensor is smoothed via removing the fluctuations. Therefore, a moving average filter is employed as defined by \cite{oppenheim1999discrete}
\begin{equation}
y_{i,j}[n] = \dfrac{1}{L} \sum_{k = 0}^{L} C_{i,j}[n-k],
\end{equation}
where $C_{i,j}[n]$ is the measured sensor voltage in the $i^{th}$ row and $j^{th}$ column of the SN, $y_{i,j}[n]$, and $L$ are the output, and window size of the filter, respectively. 

When there is no transmission from the TX, the sensors still output a positive voltage value which is defined as the offset level, i.e., $ \rho_{o_{i,j}} $. Since the offset levels of the sensors can be different, the threshold voltage ($\gamma_{i,j}$) is defined for each sensor node by employing a constant detection threshold amplitude ($A_T$) as given by
\begin{equation}
\gamma_{i,j} = \rho_{o_{i,j}} + A_T.
\end{equation}
In order to calculate $ \rho_{o_{i,j}} $, the first $p$ samples of the received signal is averaged before the moving average filter. After $\gamma_{i,j}$ is determined for each sensor node, the amplitude detection is made by
\begin{equation}
	y_{i,j}[n] \overunderset{\mathcal{H}_1}{\mathcal{H}_0}{\gtrless}\gamma_{i,j},
	\label{amp_detection}
\end{equation}
$\mathcal{H}_1$ and $\mathcal{H}_0$ represent the hypotheses when the signal is present or absent, respectively. The time instances that each sensor reaches the $\gamma_{i,j}$ value in $y_{i,j}[n]$ is recorded as $t_{i,j}$. Next, the energy detection scheme is explained.

\subsubsection{Energy Detection Scheme}
In the amplitude detection scheme, the detection time is determined according to a voltage threshold after a moving average filter. However, the energy detection scheme is implemented by only using an energy threshold without a filter. 

\begin{figure}[tb]
	\centering
	\includegraphics[width=2.7in]{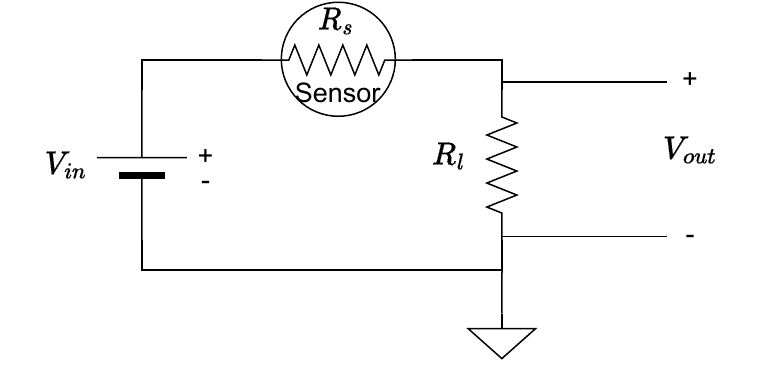}
	\caption{Measurement circuit of the sensor board.}
	\label{Circuit} \vspace{-0.3cm}
\end{figure}

As applied in the amplitude detection scheme, $ \rho_{o_{i,j}} $ is calculated by averaging the first $p$ samples of the received signals. Because the offset levels of the sensors can be different, the offset is subtracted from the received signals before the energy detection which is made according to 
\begin{equation}
	\frac{1}{R_l}\sum_{n=1}^{N_s} g_{i,j}[n]^2 \overunderset{\mathcal{H}_1}{\mathcal{H}_0}{\gtrless}\lambda,
	\label{detection_rule}
\end{equation}
where $N_s$ is the number of samples for energy detection, $\lambda$ is the energy threshold, $R_l$ is the load resistance of the measurement circuit, $\mathcal{H}_1$ and $\mathcal{H}_0$ represent the hypotheses when the signal is present or absent, respectively. Also, $g_{i,j}[n]$ is the non-offset measured sensor voltage calculated as $g_{i,j}[n] = V_{out} - \rho_{o_{i,j}}$. In (\ref{detection_rule}), the left-hand side of the equation gives the energy dissipated on the $R_l$ where the voltage is measured as shown in Fig. \ref{Circuit}.  The instantaneous voltage value at which the energy of each signal reaches $\lambda$ are recorded as $\gamma_{i,j}$ and the time instance as $t_{i,j}$. These  $t_{i,j}$ and $\gamma_{i,j}$ values are employed as input for velocity estimation and the sensitivity response of the sensor, respectively. A more detailed statistical analysis of the SN is given in  Section \ref{detection}.

\subsection{Sensitivity Response of the Sensor} \label{sub_sensitivity}
Sensor voltage is obtained via a sensor measurement circuit on the MQ-3 sensor boards which is shown in Fig. \ref{Circuit}. As the concentration around the sensor changes, its resistance ($R_s$) changes. Hence, the molecule concentration is converted to an electrical signal via the circuit in Fig. \ref{Circuit} where $V_{out}$ gives the output voltage. Using this circuit, $R_s$ is derived as
\begin{equation}
	R_s = \left(\frac{V_{in}}{V_{out}} - 1\right) R_l,
	\label{R_s}
\end{equation}
where $ V_{in} $ shows the DC input voltage and $R_l$ is the load resistance. 

For each concentration value, the sensor has a different $R_s$ value. The sensor resistance can be normalized by dividing $R_s$ to $R_o$, where $R_o$ is the sensor resistance measured at the concentration value of $0.0004$ kg/m$^3$, the minimum concentration level MQ-3 sensor can measure \cite{MQ3}. According to its datasheet, MQ-3 sensor has a sensitivity characteristic which maps each concentration value to the normalized resistance value ($R_s/R_o$) \cite{MQ3}. This sensitivity characteristic can be expressed as a sensitivity function
\begin{equation}
	f\left(C_{i,j}\right) = \frac{R_s}{R_o} = \left(\frac{V_{in}}{V_{out}} - 1\right) \frac{R_l}{R_o},
	\label{f1}
\end{equation}
where $C_{i,j}$ is the actual molecule concentration around the sensor and $R_s/R_o$ is given by substituting (\ref{R_s}) into (\ref{f1}). By employing the values in its datasheet, $f(C_{i,j})$ can be obtained via curve fitting technique. Nonlinear least squares method that minimizes the sum of the square errors is employed to fit the datasheet values of the MQ-3 sensitvity characteristic. As the result of the curve fitting, $f(C_{i,j})$ is given by
\begin{equation}
	f\left(C_{i,j}\right) =  a_1 \left(C_{i,j}\right)^{b_1}+ d_1,
	\label{f2}
\end{equation}
where $a_1$, $b_1$ and $d_1$ are the curve fitting parameters. By employing Levenberg-Marquardt algorithm \cite{hagan1994training}, these parameters are estimated as $a_1=0.0116$, $b=-0.5855$ and $d_1=-0.0743$ with a Root Mean Square Error (RMSE) value of $0.0371$. The Levenberg-Marquardt algorithm, which is a widely employed iterative optimization technique, effectively minimizes the sum of squared differences between the modeled and actual data, refining the alignment between \(f(C_{i,j})\) and the MQ-3 sensitivity characteristic given in its datasheet \cite{MQ3}. This algorithm dynamically adjusts the parameters \(a_1\), \(b_1\), and \(d_1\) via an iterative process, thus intricately tailoring the curve fitting to mitigate the discrepancy between the model and empirical MQ-3 sensitivity data. Specifically, the algorithm integrates local Gauss-Newton convergence and global steepest descent methods to optimize these parameters. This dynamic technique ensures accurate curve fitting across a wide spectrum of residual values, elevating the precision of the sensor model. The sensitivity response of the MQ-3 sensor is also employed in our previous study \cite{gulec2021droplet} which is a part of the signal reconstruction approach of the RX in macroscale. The signal reconstruction of the RX is first proposed in \cite{atakan2019signal} in order to investigate how the actual concentration around the RX is sensed in microscale.

In order to find the molecule concentration for the given detection threshold voltage,  $V_{out}$ is set as $\gamma_{i,j}$ in (\ref{f1}). Equations (\ref{f1}) and (\ref{f2}) are combined to obtain the equation as given by
\begin{equation}
	\left(\frac{V_{in}}{V_{out}} - 1\right) \frac{R_l}{R_o} = a_1 \left(C_{i,j}\right)^{b_1}+ d_1.
	\label{f3}
\end{equation}
As the result of the sensitivity response of the sensor, (\ref{f3}) is manipulated to obtain $C_{i,j}$,   which is given by
\begin{equation}
	C_{i,j} = \left(\frac{V_{in}R_l - \gamma_{i,j} R_l - d_1 \gamma_{i,j} R_o}{\gamma_{i,j} R_o a_1}\right)^{(1/b_1)}.
	\label{C_{i,j}}
\end{equation}
\subsection{Wind Velocity Estimation}
After the voltages ($\gamma_{i,j}$) and  detection times ($ t_{i,j} $) are obtained for each sensor, the wind velocity flowing over two sensor nodes can be estimated in $ x $ and $ y $ directions as given by
\begin{equation}
	u_x = \frac{|x_2-x_1|}{t_2-t_1}, u_y = \frac{|y_2-y_1|}{t_2-t_1},
	\label{u_xy}
\end{equation}
where ($ x_1,y_1 $) and ($ x_2,y_2 $) are the coordinates for the first and second sensor node, respectively and $t_1$ and $t_2$ are the detection times for the first and second sensor node, respectively. In our scenario, the wind velocity estimation method described in (\ref{u_xy}) is generalized by averaging the wind velocities estimated by sensor node pairs within the corresponding clusters, as defined in Fig. \ref{SN_blocks}, for four directions. The directions of these velocities in the 2D Cartesian coordinate system are indicated by the symbols $'+'$ and $'–'$. The wind velocity averages are calculated as given below
\begin{align}
	\bar{u}_{x_-} &= \frac{1}{M_r}\sum_{i=1}^{M_r} u_{x_-}^{(i)} =  \frac{1}{M_r}\sum_{i=1}^{M_r}\frac{|x_{i,1}-x_{i,2}|}{(t_{i,1}-t_{i,2})}, \ \text{Cluster-1}\\
	\bar{u}_{x_+} &= \frac{1}{M_r}\sum_{i=1}^{M_r} u_{x_+}^{(i)} = \frac{1}{M_r}\sum_{i=1}^{M_r}\frac{|x_{i,5}-x_{i,4}|}{(t_{i,5}-t_{i,4})},\ \text{Cluster-3}\\
	\bar{u}_{y_+} &= \frac{1}{M_c}\sum_{j=1}^{M_c} u_{y_+}^{(j)} = \frac{1}{M_c}\sum_{j=1}^{M_c}\frac{|y_{5,j}-y_{4,j}|}{(t_{5,j}-t_{4,j})},\ \text{Cluster-2}\\
	\bar{u}_{y_-} &= \frac{1}{M_c}\sum_{j=1}^{M_c} u_{y_-}^{(j)} = \frac{1}{M_c}\sum_{j=1}^{M_c}\frac{|y_{1,j}-y_{2,j}|}{(t_{1,j}-t_{2,j})}, \text{ Cluster-4}
\end{align}
where $u_{x_\pm}^{(i)}$ and $u_{y_\pm}^{(j)}$ show the instantaneous wind velocity of the $i^{th}$ and $j^{th}$ sensor node pair in the corresponding direction (or cluster), respectively, $\bar{u}_{x_\pm}$ and $\bar{u}_{y_\pm}$ represent the averages of velocities estimated by these sensor node pairs in the corresponding direction (or cluster), respectively, $x_{i,j}$ and $y_{i,j}$ indicate the horizontal and vertical position of the sensor node $N_{i,j}$ given in Fig. \ref{SN_diagram}, respectively, and $M_r$ and $M_c$ are the total number of rows and columns of the SN, respectively. Here, the instantaneous velocities whose values are negative are not considered for the velocity estimation. During the experiments, it is observed that the wind blows stronger in one direction which means that it can only have at most two velocity components among the estimated velocities in four directions. Therefore, $u_x$ and $u_y$ are defined as
\begin{equation}
	u_x = \text{max}(\bar{u}_{x_-},\bar{u}_{x_+}), \quad u_y = \text{max}(\bar{u}_{y_-},\bar{u}_{y_+}).
\end{equation}
\subsection{Transmitted Mass Calculation}
The estimated values of $u_x$ and $u_y$ are used to calculate the evaporation rate of ethanol in the air ($Q_e$). For the wind blowing over a surface with a velocity $u$ at room temperature ($25^\circ C$), $Q_e$ (kg/m$^2$s) is given by \cite{lyulin2015measuring}
\begin{equation}
	Q_e = h_1 u^{0.54},
\end{equation}
where $u=\sqrt{u_x^2 + u_y^2}$ and $h_1=4\times 10^{-3}$ $kg/m^3$, which is the constant for ethanol as given in \cite{lyulin2015measuring}. In order to find the mass flow rate of evaporated molecules, i.e., $Q$ (kg/s), which is defined as the mass flowing through a surface per unit time, $Q = Q_e A$ where $A$ is the surface area of evaporated molecules, i.e., the surface area of the petri dish for our case. Here, we assume that the source with a short emission time value ($T_e$) is considered as an instantaneous source. Hence, the transmitted mass can be calculated as \cite{munson2009brief}
\begin{equation}
	m_T = Q T_e = Q_e A T_e.
	\label{m_T}
\end{equation}
\subsection{Operation of the SNCLA} 
Thus far, the required input parameters for the location estimator (see Fig. \ref{SN_blocks}) are obtained. By using these parameters, Algorithm \ref{Alg1} is proposed for the localization of the TX. In this algorithm, two clusters are chosen according to the direction of the wind velocity on $x$ and $y$ axes. For instance, if the wind blows stronger in the $+x$ direction on the $x$-axis and $+y$ direction on the $y$ axis, then the node pairs in Cluster-$  3 $ and Cluster-$ 2 $ are chosen for the location estimation. Similar to the wind velocity estimation, the node pairs whose absolute instantaneous wind velocity value is negative are not considered for the location estimation. Afterwards, the equation pairs given in  (\ref{Clus1})-(\ref{Clus4}) are solved for $x_T$ and $y_T$ according to the chosen two clusters. The solution for each of two equations gives the estimated coordinates of the TX, i.e., $\hat{x}_T$ and $\hat{y}_T$. Equations (\ref{Clus1})-(\ref{Clus4}) are solved numerically, as given in the numerical results in the next section. 

\begin{algorithm}[tb]
	\caption{SNCLA}
	\label{Alg1} 
	\begin{algorithmic}
		\STATE \textbf{input:} $ \bar{u}_{x_\pm}$, $ \bar{u}_{y_\pm}$, $u_{x_\pm}^{(i)}$, $u_{y_\pm}^{(j)}$, $m_T$, $C_{i,j}$, $t_{i,j}$ for all \\$i = 1,...,M_r$, $j=1,...,M_c$
	\IF{($ u_x == u_{x_-} $) and ($ u_y == u_{y_+} $)}
	\STATE Calculate ($\hat{x_T},\hat{y_T}$) by (\ref{Clus1}) for Cluster 1  \STATE Calculate ($\hat{x_T},\hat{y_T}$) by (\ref{Clus2}) for Cluster 2
	\ELSIF{($ u_x == u_{x_-}$) and ($ u_{y_-} == u_{y_+}$)}
	\STATE Calculate ($\hat{x_T},\hat{y_T}$) by (\ref{Clus1}) for Cluster 1  \STATE Calculate ($\hat{x_T},\hat{y_T}$) by (\ref{Clus4}) for Cluster 4
	\ELSIF{($ u_x == u_{x_+}$) and ($ u_y == u_{y_+}$)}
	\STATE Calculate ($\hat{x_T},\hat{y_T}$) by (\ref{Clus2}) for Cluster 2  \STATE Calculate ($\hat{x_T},\hat{y_T}$) by (\ref{Clus3}) for Cluster 3
	\ELSE
	\STATE Calculate ($\hat{x_T},\hat{y_T}$) by (\ref{Clus3}) for Cluster 3  \STATE Calculate ($\hat{x_T},\hat{y_T}$) by (\ref{Clus4}) for Cluster 4
	\ENDIF 
	\end{algorithmic}
\end{algorithm}
	
	\begin{subequations}
	\begin{align}
		\frac{(x_{i,1}\hspace{-0.1cm} -\hspace{-0.1cm} x_T\hspace{-0.1cm} -\hspace{-0.1cm} u_x t_{i,1})^2}{2\sigma_x^2}\hspace{-0.1cm} +\hspace{-0.1cm} \frac{(y_{i,1}\hspace{-0.1cm} -\hspace{-0.1cm} y_T\hspace{-0.1cm} -\hspace{-0.1cm} u_y t_{i,1})^2}{2\sigma_y^2}\hspace{-0.1cm} +\hspace{-0.1cm} n_{i,1} &= 0  \\ \frac{(x_{i,2}\hspace{-0.1cm} -\hspace{-0.1cm} x_T\hspace{-0.1cm} -\hspace{-0.1cm} u_y t_{i,2})^2}{2\sigma_x^2}\hspace{-0.1cm}+\hspace{-0.1cm} \frac{(y_{i,2}\hspace{-0.1cm} -\hspace{-0.1cm} y_T\hspace{-0.1cm} -\hspace{-0.1cm} u_y t_{i,2})^2}{2\sigma_y^2}\hspace{-0.1cm} +\hspace{-0.1cm} n_{i,2} &= 0 
	\end{align}
\label{Clus1}
\end{subequations}
\begin{subequations}
	\begin{align}
        \frac{(x_{5,j}\hspace{-0.1cm} -\hspace{-0.1cm} x_T\hspace{-0.1cm} -\hspace{-0.1cm} u_x t_{5,j})^2}{2\sigma_x^2}\hspace{-0.1cm} +\hspace{-0.1cm} \frac{(y_{5,j}\hspace{-0.1cm} -\hspace{-0.1cm} y_T\hspace{-0.1cm} -\hspace{-0.1cm} u_y t_{5,j})^2}{2\sigma_y^2}\hspace{-0.1cm} +\hspace{-0.1cm} n_{5,j} &= 0 \\ \frac{(x_{4,j}\hspace{-0.1cm} -\hspace{-0.1cm} x_T\hspace{-0.1cm} -\hspace{-0.1cm} u_x t_{4,j})^2}{2\sigma_x^2}\hspace{-0.1cm} +\hspace{-0.1cm} \frac{(y_{4,j}\hspace{-0.1cm} -\hspace{-0.1cm} y_T\hspace{-0.1cm} -\hspace{-0.1cm} u_y t_{4,j})^2}{2\sigma_y^2}\hspace{-0.1cm} +\hspace{-0.1cm} n_{4,j} &= 0 
	\end{align}
\label{Clus2}
\end{subequations}
\begin{subequations}
	\begin{align}
		\frac{(x_{i,5}\hspace{-0.1cm} -\hspace{-0.1cm} x_T\hspace{-0.1cm} -\hspace{-0.1cm} u_x t_{i,5})^2}{2\sigma_x^2}\hspace{-0.1cm} +\hspace{-0.1cm} \frac{(y_{i,5}\hspace{-0.1cm} -\hspace{-0.1cm} y_T\hspace{-0.1cm} -\hspace{-0.1cm} u_y t_{i,5})^2}{2\sigma_y^2}\hspace{-0.1cm} +\hspace{-0.1cm} n_{i,5} &= 0 \\ \frac{(x_{i,4}\hspace{-0.1cm} -\hspace{-0.1cm} x_T\hspace{-0.1cm} -\hspace{-0.1cm} u_y t_{i,4})^2}{2\sigma_x^2}\hspace{-0.1cm} +\hspace{-0.1cm} \frac{(y_{i,4}\hspace{-0.1cm} -\hspace{-0.1cm} y_T\hspace{-0.1cm} -\hspace{-0.1cm} u_y t_{i,4})^2}{2\sigma_y^2}\hspace{-0.1cm} +\hspace{-0.1cm} n_{i,4} &= 0 
	\end{align}
	\label{Clus3}
\end{subequations}
\begin{subequations}
	\begin{align}
		\frac{(x_{1,j}\hspace{-0.1cm} -\hspace{-0.1cm} x_T\hspace{-0.1cm} -\hspace{-0.1cm} u_x t_{1,j})^2}{2\sigma_x^2}\hspace{-0.1cm} +\hspace{-0.1cm} \frac{(y_{1,j}\hspace{-0.1cm} -\hspace{-0.1cm} y_T\hspace{-0.1cm} -\hspace{-0.1cm} u_y t_{1,j})^2}{2\sigma_y^2}\hspace{-0.1cm} +\hspace{-0.1cm} n_{1,j} &= 0 \\ \frac{(x_{2,j}\hspace{-0.1cm} -\hspace{-0.1cm} x_T \hspace{-0.1cm}-\hspace{-0.1cm} u_x t_{2,j})^2}{2\sigma_x^2}\hspace{-0.1cm} +\hspace{-0.1cm} \frac{(y_{2,j}\hspace{-0.1cm} -\hspace{-0.1cm} y_T \hspace{-0.1cm}-\hspace{-0.1cm} u_y t_{2,j})^2}{2\sigma_y^2}\hspace{-0.1cm} +\hspace{-0.1cm} n_{2,j} &= 0 
	\end{align}
	\label{Clus4}
\end{subequations}

\subsection{Numerical Results}
\label{Results}
In this section, numerical results of the SNCLA is given. $25$ measurements each lasting $ 180 $ s were performed with the experimental platform. There were at least $30$ min left among adjacent measurements in order to decrease the concentration level with the ventilation of the fume hood. The ventilation was not used during the measurements.

\begin{table}[!b]
	\caption{Experimental parameters.\label{tab:table1}}
	\centering
	\begin{tabular}{p{140pt}|p{80pt}}
		\hline \hline
		\textbf{Parameter}	& \textbf{Value}\\
		\hline \hline
		Number of measurements ($M_m$) & $ 25 $\\
		\hline
		Energy threshold  ($\lambda$) & $ 4.3 $ mJ\\
		\hline
		Emission time ($T_e$) & 0.1 s\\
		\hline
		Actual TX location $(x_T,y_T)$ & ($ 0.3,0.3 $) m\\
		\hline
		Area of the petri dish ($A$) & 0.0024 m$ ^2 $  \\
		\hline
		Input voltage of the sensor board ($V_{in}$) & $5$ V \\
		\hline
		Load resistance ($R_l$) & $1$ k$\Omega$\\
		\hline
		Sensor resistance at $0.0004$ kg/m$ ^3 $ ($R_o$) & $24$ k$\Omega$ \\
		\hline
		Standard deviation of the Gaussian concentration distribution on the x, y and z axis ($\sigma_x, \sigma_y, \sigma_z$) & $ 0.0115 $ m, $ 0.0115 $ m, $ 0.0046 $ m\\
		\hline
        Window size of the moving average \\ filter for amplitude detection (\(L\)) & 7 \\ \hline
		Number of samples to be averaged \\for the offset level $ \rho_{o_{i,j}} $ of the sensors  ($p$) & 50\\ 
		\hline \hline
	\end{tabular}
\end{table}

\begin{figure}[tb]
	\centering
	\includegraphics[width=\columnwidth]{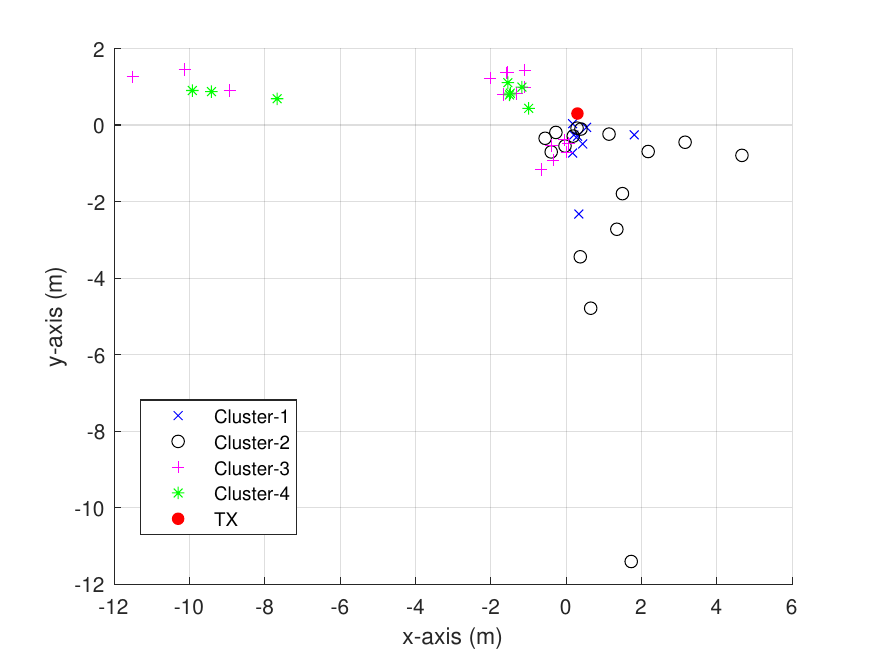}
	\caption{Estimated points using SNCLA for all clusters.}
	\label{Plot_cluster_all}
\end{figure}

\begin{figure}[!b]
	\centering
	\includegraphics[width=\columnwidth]{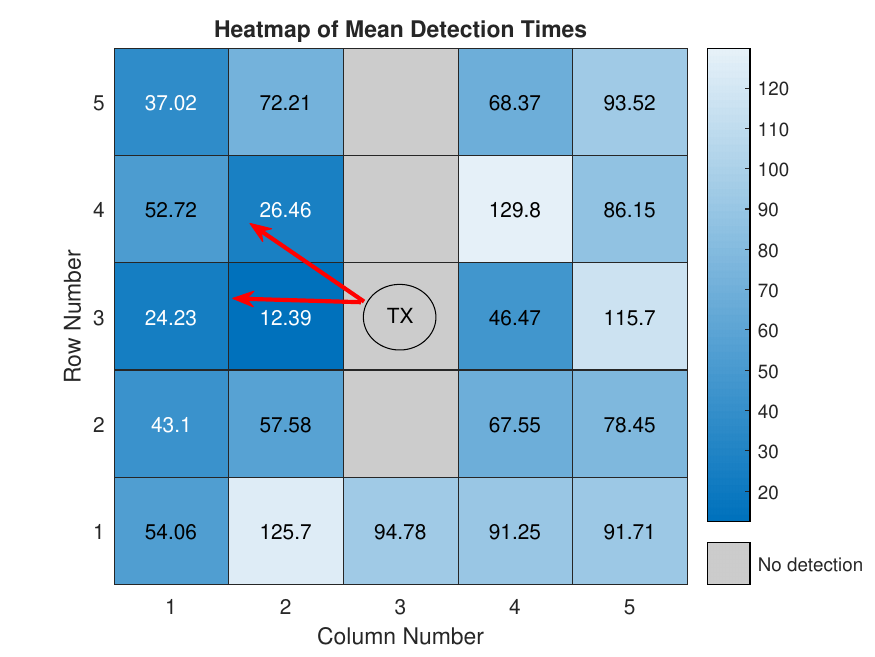}
	\caption{Mean detection times (s) for each sensor node. The source is deployed in the center and the dominant wind directions are shown with the arrows.}
	\label{Plot_heatmap_time}
\end{figure}

\begin{figure}[!t]
	\centering
	\includegraphics[width=\columnwidth]{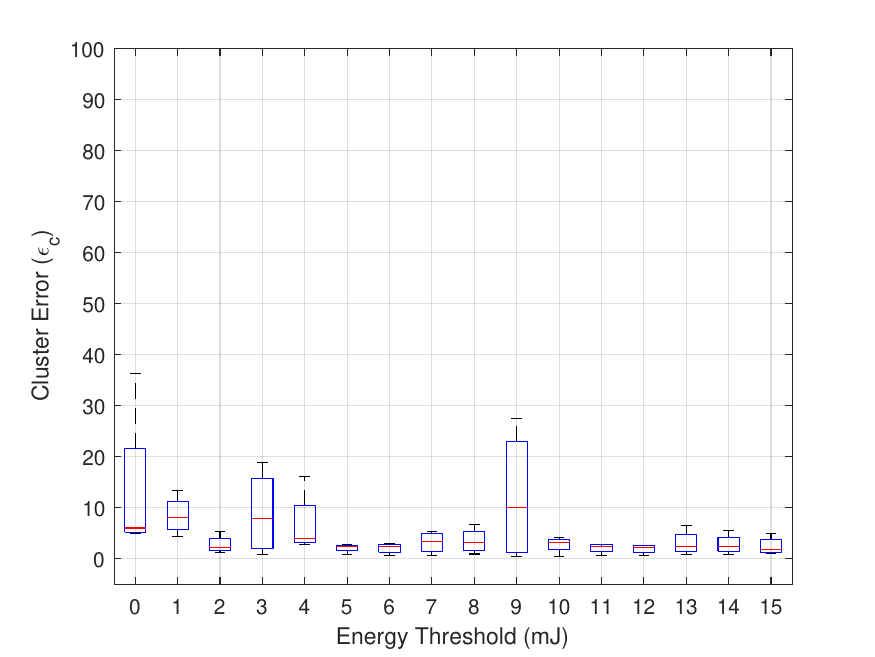}%
	\caption{Box plot of cluster error vs. threshold for energy detection scheme.}
		\label{fig_first_case}
\end{figure}

The experimental parameters are given in Table \ref{tab:table1}. Among these parameters, $R_o$ is calculated
by employing output voltage ($V_{out}$) of the sensor measurement circuit. According to the MQ-3 sensor datasheet, its detection scope is between  $ 5 \times 10^{-5} $ and $10^{-2}$ kg/m$ ^3 $ \cite{MQ3}. This detection scope is scaled for $V_{out}$ values between $0$ and $5$ V. Thus, $0.0004$ kg/m$ ^3 $ corresponds to $V_{out}=0.2$ V which is used to calculate the sensor resistance value via (\ref{R_s}). As mentioned in Section \ref{System_model}, the dispersion parameters ($\sigma_x, \sigma_y, \sigma_z$) are assumed as constant values. For our experimental scenario and according to (\ref{sigma_y})-(\ref{sigma_z}), the ranges of $\sigma_y$ and $\sigma_z$ are $ 0.006-0.017 $ m and $0.0024-0.0068 $ m. Therefore, $ \sigma_y $ and $\sigma_z$ are chosen as the average values of these ranges. $\sigma_x$ is also taken as equal to $\sigma_y$ \cite{de2013air}. In addition, $\lambda$ is also chosen as an empirical value in order to have accurate estimations as given in Fig. \ref{Plot_cluster_all} and \ref{Plot_heatmap_time}. However, the error values for a range of $\lambda$ values are also given in Fig. \ref{fig_first_case}.

The estimation results for each cluster are given in Fig. \ref{Plot_cluster_all}. For the experimental values, $\hat{x}_T$ and $\hat{y}_T$ have two complex conjugate roots for each. Therefore, only real parts of the solutions are considered for the numerical results. As shown in Fig. \ref{Plot_cluster_all}, the wind speed for our measurements is mostly in the $-x$ and $+y$ directions. The SNCLA gives better results when the wind speed is higher, as the effect of dispersion of evaporating molecules is reduced. Therefore, Cluster-$1$ and Cluster-$2$ results are also more accurate than other clusters.

In Fig. \ref{Plot_heatmap_time}, a heatmap illustrates the average detection times across each sensor node, accompanied by directional arrows that highlight the prevailing wind's orientation. Notably, the third column of the SN reveals a solitary sensor detection pattern. This observation bears that certain evaporating molecules align their motion with the wind's trajectory, indicated by concurrent movement, while others predominantly exhibit horizontal motion opposite to the wind. This behavior can be attributed to the initial puff of the TX, which influences the movement of molecules, underscoring the complexity of the diffusion process.

%In Fig. \ref{Plot_heatmap_time}, the average of the detection times for each sensor node is given as a heatmap. Mean detection times verify the direction of the wind as also shown with the arrows in this figure. In the third column of the SN there is only one sensor detection. According to this observation, some of the evaporating molecules move in the same direction as the wind, while the remaining molecules mostly move in the opposite horizontal direction of the wind due to the first puff of the TX. Compared to the heatmap in \cite{gulec2020localization}, the detection time appears to be longer. This shows that while the energy detection scheme takes longer time than amplitude detection, it improves the detection accuracy as also shown in Figs. \ref{fig_first_case} and \ref{fig_second_case}.

In order to evaluate the estimation performance, the cluster error ($ \epsilon_c $), is defined for each cluster as given by
\begin{numcases}
	{\epsilon_c = } \frac{1}{M_r}\sum_{i=1}^{M_r}\frac{1}{M_m}\sum_{k=1}^{M_m}\sqrt{(x_T - \hat{x}^{(i)}_{T_{k}})^2 + (y_T - \hat{y}^{(i)}_{T_k})^2},  \\
	\frac{1}{M_c}\sum_{j=1}^{M_c}\frac{1}{M_m}\sum_{k=1}^{M_m}\sqrt{(x_T - \hat{x}^{(j)}_{T_{k}})^2 + (y_T - \hat{y}^{(j)}_{T_{k}})^2}, 
\end{numcases}
where $M_m$ is the number of the measurements and ($ \hat{x}^{(i)}_{T_{k}}, \hat{y}^{(i)}_{T_{k}} $) show the estimated points for the $i^{th}$ node pair in the corresponding cluster at the $ k^{th}$ measurement. In our case, there are $M_r = M_c = 5$ node pairs for each cluster. First, the Euclidean distance between the actual and estimated points are calculated for each node pair in the cluster. Then, these distances for all the measurements are averaged. This process is repeated for each node pair in the cluster. $\epsilon_c$ serves as an analogous yardstick to familiar performance metrics like mean square error or Euclidean distance. It quantifies the disparity between actual and estimated positions within sensor node pairs across clusters. Just as traditional metrics measure estimation accuracy, $\epsilon_c$ assesses the precision of our method in diverse cluster scenarios. This provides tailored insights into positioning quality for distinct cluster arrangements, enhancing the granularity of our evaluation.

\begin{figure}[tp]
	\includegraphics[width=\columnwidth]{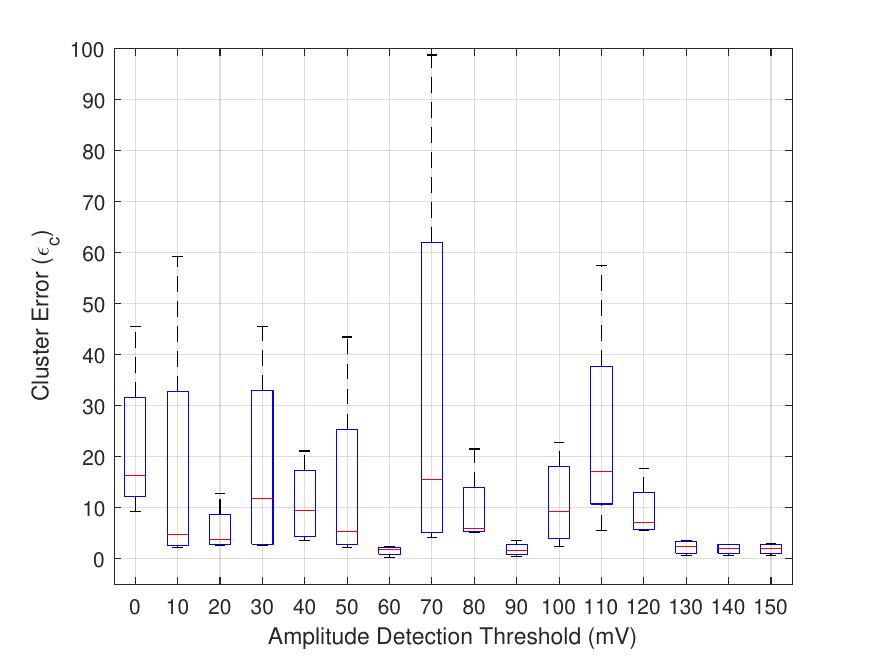}%
	\caption{Box plot of cluster error vs. threshold for amplitude detection scheme.}
	\label{fig_second_case}
\end{figure}

In Fig. \ref{fig_first_case}, $\epsilon_c$ results are given for values of $\lambda$ between $0-15$ mJ in $1$ mJ steps, and in Fig. \ref{fig_second_case} for amplitude detection threshold  values between $0-0.15$ V with $10$ mV steps. Box plot is used to visualize the distribution of $\epsilon_c$ values according to clusters. The middle mark in the box shows the median, the lower and upper edge of the box indicate the $25^{th}$ and $75^{th}$ percentile, respectively. It is observed that energy detection scheme outperforms amplitude detection scheme in terms of median accuracies and their fluctuations. Since experiments are conducted in a fume hood, there are reflected signals in the measured signals. While this deteriorates the estimation results in amplitude detection, energy detection scheme benefits from it with a cost of later detection time.

\section{Statistical Analysis}
\label{detection}
As mentioned in Section \ref{SPD}, there is a fluctutation stemming from the noise on the measured signals. This noise signal represents the combined effect of the turbulent diffusion and electronic noise dependent on the components such as cables, resistors and transistors in the experimental setup. In the previous section, our results are based on mitigating the effect of the noise on the measured signals for both amplitude detection and energy detection schemes. However, it is essential to characterize the noise on the measured signals. To this end, a statistical analysis of the received signals via the sensor network is performed in this section. First, the detection of the sensor nodes is modeled by binary hypothesis testing. Next, measured signals are statistically characterized.

The detection can be expressed by using a binary hypothesis testing with hypotheses ${\mathcal{H}}_{0}$ and ${\mathcal{H}}_{1}$ which correspond to the absence and presence of the signal, respectively.
%\vspace{-0.5 cm}
\begin{align}
	\hspace{-0.2cm}{\mathcal{H}}_{0}\colon z_{i,j}[n] &= w_{i,j}[n] + \rho_{o_{i,j}},\quad n=1,2,\ldots, N_s \label{H_0}\\
	\hspace{-0.2cm}{\mathcal{H}}_{1}\colon z_{i,j}[n] &= x_{i,j}[n] + w_{i,j}[n]+\rho_{o_{i,j}}, n=1,2,\ldots, N_s \label{H_1}
\end{align}
where subscripts $i$ and $j$ show the row and column number of the sensor node $N_{i,j}$, respectively, $w_{i,j}[n]$ denotes the noise samples, $ \rho_{o_{i,j}} $ is the constant offset signal, $x_{i,j}[n]$ is the sensed signal by the sensor, $z_{i,j}[n]$ shows the received signal and $N_s$ is the number of samples. Here, the noise signal is modeled as an additive component as used in the detection of SNs \cite{li2007distributed} and MC systems \cite{bhattacharjee2022digital}.

%By using ${\mathcal{H}}_{1}$ and ${\mathcal{H}}_{0}$, the probability of missed detection ($P_{md}$) can be defined as $P_{md} = P({\mathcal{H}}_{0}|{\mathcal{H}}_{1})$ which corresponds to the probability that the sensors detect only noise while there is alcohol in the environment. The detection probability is expressed as $P_{d} = P({\mathcal{H}}_{1}|{\mathcal{H}}_{1}) = 1-P_{md}$. In addition, the false alarm probability is given as $P_{fa} = P({\mathcal{H}}_{1}|{\mathcal{H}}_{0})$. By using (\ref{hypothysis}), the energy-based decision rule of a sensor node in the sensor network is defined as
%\begin{equation}
%		\mathbb{T}_{i,j}(\mathbf{z}_{i,j})=\sum_{n=1}^{N_s}{|z_{i,j}}[n]{{|}^{2}}\overunderset{\mathcal{H}_1}{\mathcal{H}_0}{\gtrless}\lambda\,
%	\label{decision_rule}	
%\end{equation}
%where $\lambda$ is the threshold, $\mathbb{T}_{i,j}(\mathbf{z}_{i,j})$ is the test statistic defined for the signal vector $\mathbf{z}_{i,j} = z_{i,j}[1],..., z_{i,j}[N_s]$.

\begin{figure}[tb]
	\centering
		\includegraphics[width=\columnwidth]{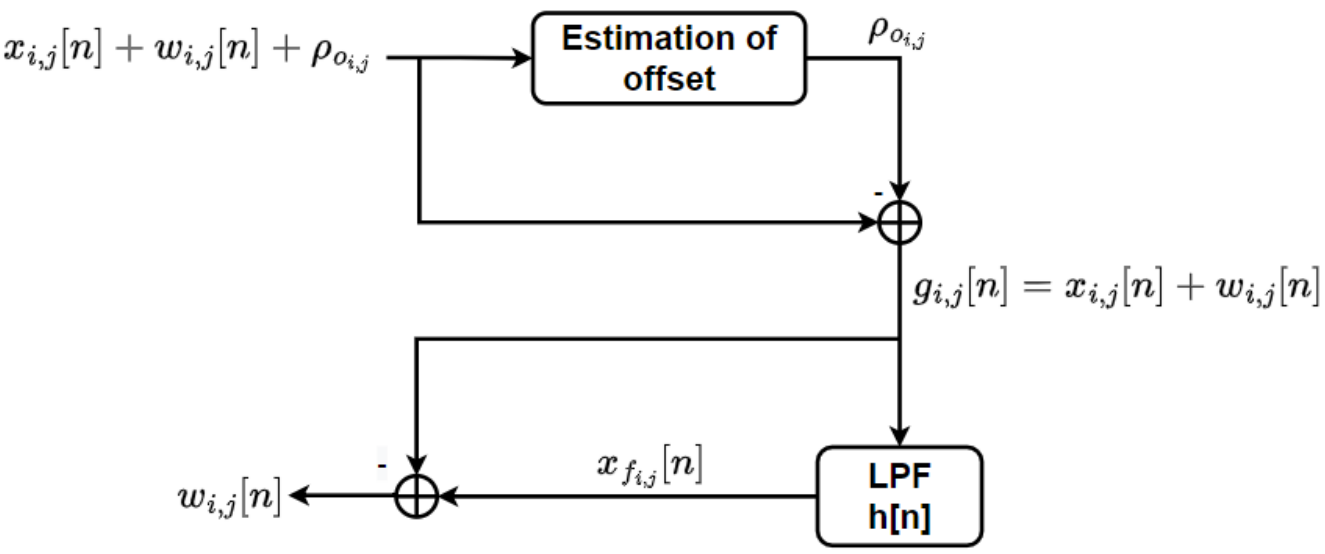}
		\caption{Block diagram of the noise extraction from the received signal.}	
		\label{block_dia}
\end{figure}

For a statistical analysis based on the energy detection rule in (\ref{detection_rule}), the statistical characteristics of the received and the noise signals need to be determined. To this end, we utilize our experimental data to estimate these characteristics. For this purpose, it is necessary to separate the noise in the received signal. The steps of this separation are depicted in the block diagram given in Fig. \ref{block_dia}. First, the average of the first $p$ samples of each signal is considered as the offset and by subtracting the offset from the received signal $\left(z_{i,j}[n]\right)$, the non-offset signal $ \left(g_{i,j}[n]\right) $ is obtained. In this paper, $p$ is empirically determined as $50$ as also used for the results in Section \ref{Results}. Then, the resulting non-offset signal is passed through the low-pass filter (LPF) and the filtered signal $\left(x_{f_{i,j}}[n]\right)$ is obtained. Since the additive noise in the received signal can be considered to be removed by the LPF, we can assume that $x[n] \approx x_f[n]$. Based on this assumption, the noise is obtained by subtracting the filtered signal from the non-offset signal as shown in Fig. \ref{block_dia}.

\begin{figure}[bt]
	\centering
	\includegraphics[width=\columnwidth]{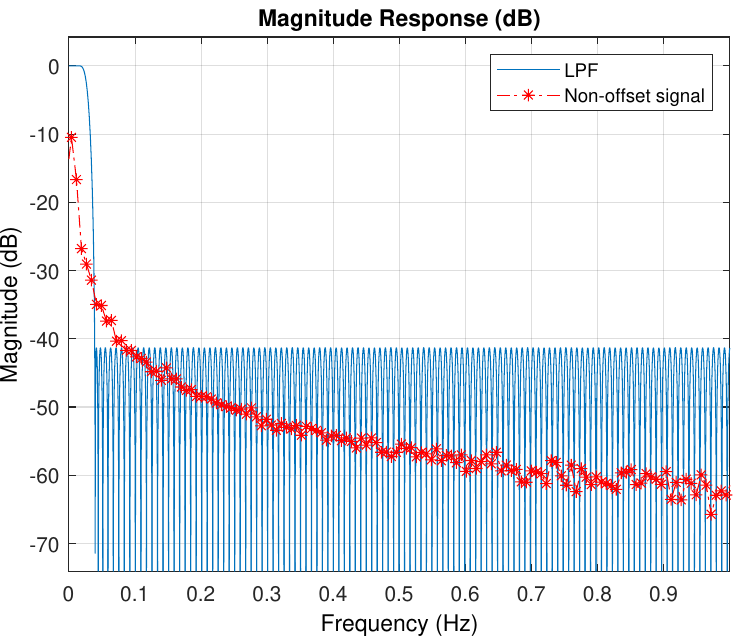}
	\caption{The frequency domain magnitude responses of the LPF and one measurement for $N_{5,5}$.}	
	\label{frequencydomain}
\end{figure}

For the noise extraction given in Fig. \ref{block_dia}, the LPF is designed as a Parks-McClellan optimal finite impulse response (FIR) filter. In the filtering process, firstly, the order of the filter is designed by using the algorithm in \cite{rabiner1973predictability}. In this algorithm, an infinite number of optimal LPFs are obtained with new passband and stopband cut-off frequencies by scaling the set of extra ripple filters. In order to determine the cut-off frequencies of the filter, the non-offset signals are analyzed in the frequency domain. As an example, the plot of a non-offset signal in the frequency domain and filter's magnitude response are given in Fig.\ref{frequencydomain}. 
	
Thus, the cut-off frequencies are chosen very close to zero so that almost all of the noise components are filtered out. Therefore, the passband and stopband cut-off frequencies are chosen as $ 0.04 $ Hz and $ 0.09 $ Hz, respectively. Secondly, a linear phase FIR filter of order $242$ is designed using the Parks-McClellan algorithm which employs the Chebyshev approximation to minimize the error in the passband and stopband, and the Remez exchange algorithm to provide the desired filter in optimal conditions \cite{digital1979ieee}. The filter's input $\left(G_{i,j}(z)\right)$ and output $\left(X_{f_{i,j}}(z) \right)$ signals in z-domain are related by \cite{parks1987digital}
\begin{equation}
	\begin{split}
	\hspace{-0.3cm}	X_{f_{i,j}}(z) &=  H(z)G_{i,j}(z)\\
		&= (h(1)+h(2)+\cdots+ h(n+1)z^{-n})G_{i,j}(z),
	\end{split}
\end{equation}
where $H(z)$ is the LPF's impulse response, $h(1), h(2),..., h(n+1)$ are the polynomial coefficients of the LPF. The noise signal is obtained after subtracting the non-offset signal from the filtered signal.

\begin{figure}[tb]
	\centering
	\includegraphics[width=\columnwidth]{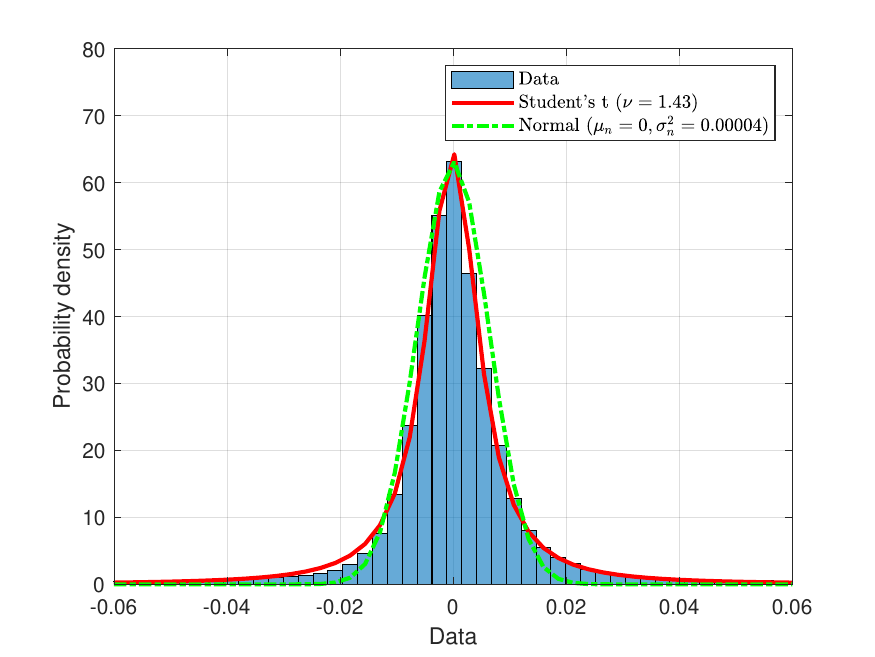}
	\caption{Histogram of the noise samples and the fitted Student's t-distribution ($\nu = 1.43$, Mean Square Error $(MSE) = 0.0912$) and normal distribution ($\mu_n = 0, \sigma_n^2 = 0.00004$, $MSE = 0.9086$).}
	\label{noise_dist}
\end{figure}

Afterwards, the histograms of $x_{i,j}[n]$ and  $w_{i,j}[n]$ are obtained by using the signal samples and fitted by using the Levenberg-Marquardt algorithm \cite{hagan1994training}, which is the same method employed for the sensitivity response of the sensor explained in Section \ref{sub_sensitivity}, according to different distributions detailed as follows. As given in Fig. \ref{noise_dist}, the noise samples  are fitted with the given parameters in the caption of the figure according to the probability density function (pdf) of the Student's t-distribution which is given by \cite{montgomery2010applied}
\begin{equation}
	\begin{split}
		f_s(\alpha)=\frac{\Gamma\left(\frac{\nu+1}{2}\right)}{\sqrt{\nu \pi} \Gamma\left(\frac{\nu}{2}\right)}\left(1+\frac{\alpha^{2}}{\nu}\right)^{-\left(\frac{\nu+1}{2}\right)},
	\end{split}
\end{equation}
where $\nu$ is the degree of freedom and $\Gamma(\alpha) = \int_{0}^{\infty} s^{\alpha-1}e^{s} \,ds$ is the gamma function. In the literature, the noise is assumed to have a Gaussian distribution for sensor networks \cite{li2007distributed,atapattu2014energy} and for practical MC systems \cite{bhattacharjee2022digital} to represent the random effects in the channel. However, it is revealed that this assumption does not hold for our scenario as shown with the Student's t-distribution in Fig. \ref{noise_dist}, although it has a similar characteristic with the Gaussian distribution having a zero mean and a symmetrical curve shape, which is also shown in Fig. \ref{noise_dist}. Normally, a Student's t-distribution can be approximated to a Gaussian distribution as $\nu$ approaches infinity according to the Central Limit Theorem \cite{montgomery2010applied}. Typically, when $\nu > 30$, it can be assumed as a Gaussian distribution. However, our case $\nu = 1.43$ is not appropriate for this approximation.

\begin{figure}[bt]
	\centering
	\includegraphics[width=\columnwidth]{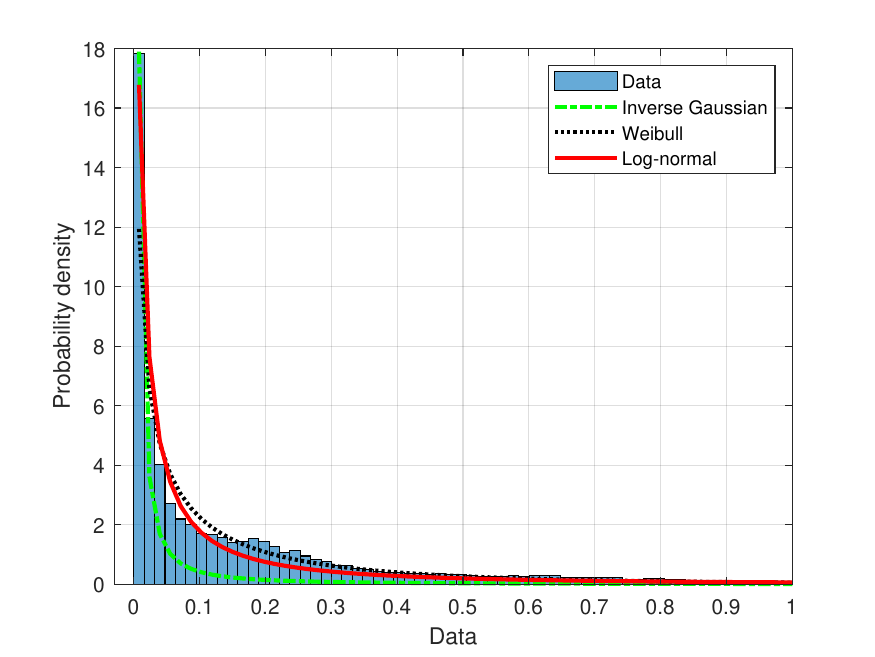}
	\caption{Histogram of $x_{f_{i,j}}[n]$ samples and its fitted inverse Gaussian, Weibull, and log-normal distributions. See Table \ref{Fit_param} for details.}
	\label{signal_dist}
\end{figure}

In addition, received signals, each lasting 180 seconds, cause an error in the filtered signal distribution due to the reflected signals in the fume hood. Therefore, the first 90 seconds of signals are used to obtain the histogram of the filtered signal as shown in Fig. \ref{signal_dist}. Here, the histogram is fitted according to three different long-tailed distributions such as inverse Gaussian, Weibull, and log-normal distributions as shown in Fig. \ref{signal_dist}. The pdfs, MSEs, fitted parameters of these distributions are given in Table \ref{Fit_param}. Although all of them have similar fits to the observed histogram, the best fit is accomplished by the log-normal distribution whose pdf is given by \cite{montgomery2010applied}
\begin{equation}
	f_{ln}(\alpha) = \frac{1}{\alpha \sigma_{\ln} \sqrt{2\pi}} \exp\left(-\frac{(\ln(\alpha) - \mu_{\ln})^2}{2 \sigma_{\ln}^2}\right),
\end{equation}
where $\mu_{\ln}$ is the location parameter, and $\sigma_{\ln}$ is the scale parameter of the distribution. Here, the log-normal distribution shows that the signal $x_{f_{i,j}}[n]$ can be interpreted as a random variable as a log-transformed version of a normally distributed random variable. This long tailed distribution given in Fig. \ref{signal_dist} overlaps by the observations in sensor-based MC systems  due to the diffusive propagation of the dispersing molecules \cite{farsad2014channel}. Moreover, a long-tailed stable distribution similar to log-normal distribution is also observed in the sensed signals in air-based MC channels with a constant air velocity in \cite{gulec2023computational}. All these  experimental results show the difficult nature of MC channels in practical applications. Furthermore, our  statistical characterization results enable the analysis for energy detection by deriving the optimal detection thresholds and detection probability, which is planned as the future work.

\begin{table*}[tb]
	\centering
	\caption{Estimated statistical parameters of the fitted distributions in Fig. \ref{signal_dist}}
\begin{tabular}{|l|l|l|l|}
\hline
Name & Pdf & Fitted parameters & MSE \\ \hline
Inverse Gaussian & $f_{ig}(\alpha) =  \sqrt{\frac{\lambda_{ig}}{2 \pi \alpha^3}} \exp{ \left(-\lambda_{ig} \frac{(\alpha-\mu_{ig})^2}{2 \mu_{ig}^2 \alpha}\right)} $ & $\mu_{ig} = 0.1726$, $\lambda_{ig} = 0.0011$  & $0.3626$ \\ \hline
Weibull & $f_w(\alpha) =\frac{\eta}{\theta}\left(\frac{\alpha}{\theta}\right)^{\eta-1} \exp{({-\frac{\alpha}{\theta}})^\eta}$ & $\eta = 0.618$, $\theta = 0.123$ & $0.3918$ \\ \hline
Log-normal & $f_{ln}(\alpha) = \frac{1}{x \sigma_{\ln} \sqrt{2\pi}} \exp\left(-\frac{(\ln(x) - \mu_{\ln})^2}{2 \sigma_{\ln}^2}\right) $ & $\mu_{\ln} = -3.0554, \sigma_{ln} = 2.0888$  & $0.1$ \\ \hline
\end{tabular}
\label{Fit_param}
\end{table*}

%($\eta = 0.618$, $\theta = 0.123$, $MSE = 0.0536$)	
	
\section{Conclusion}
\label{Conclusion}
This paper presents a novel experimental platform for macroscale MC and a novel algorithm for the localization of a molecular TX with a sensor network of four clusters, i.e., SNCLA. In our experimental platform, the molecular TX emits molecules by evaporation at room temperature and the signals are received with the SN. First, Gaussian plume model is given as the system model for our scenario. Based on this system model, a location estimator  is derived. Then, estimation/calculation methods for the unknown parameters in the location estimator such as detection time, transmitted mass, wind velocity and the actual concentration are proposed. Finally, SNCLA is explained by combining all these estimated/calculated parameters to find the location of the TX. SNCLA gives more accurate results for the clusters in the same direction with the wind for higher detection threshold values. Since the Gaussian plume model on which the SNCLA is based is employed for longer distances in the meteorology domain, it is anticipated to have more accurate results on larger scales with the proposed SNCLA. Furthermore, the statistical analysis reveals that the additive noise and the sensed signal are characterized by Student's t and Weibull distributions, respectively. As the future work, we plan to improve this model on larger scales and adapt different localization algorithms from the sensor network literature with  MC perspective.

\bibliographystyle{IEEEtran}
\bibliography{IEEEabrv, sn_fg}

\begin{IEEEbiography}[{\includegraphics[width=1in,height=1.25in,clip,keepaspectratio]{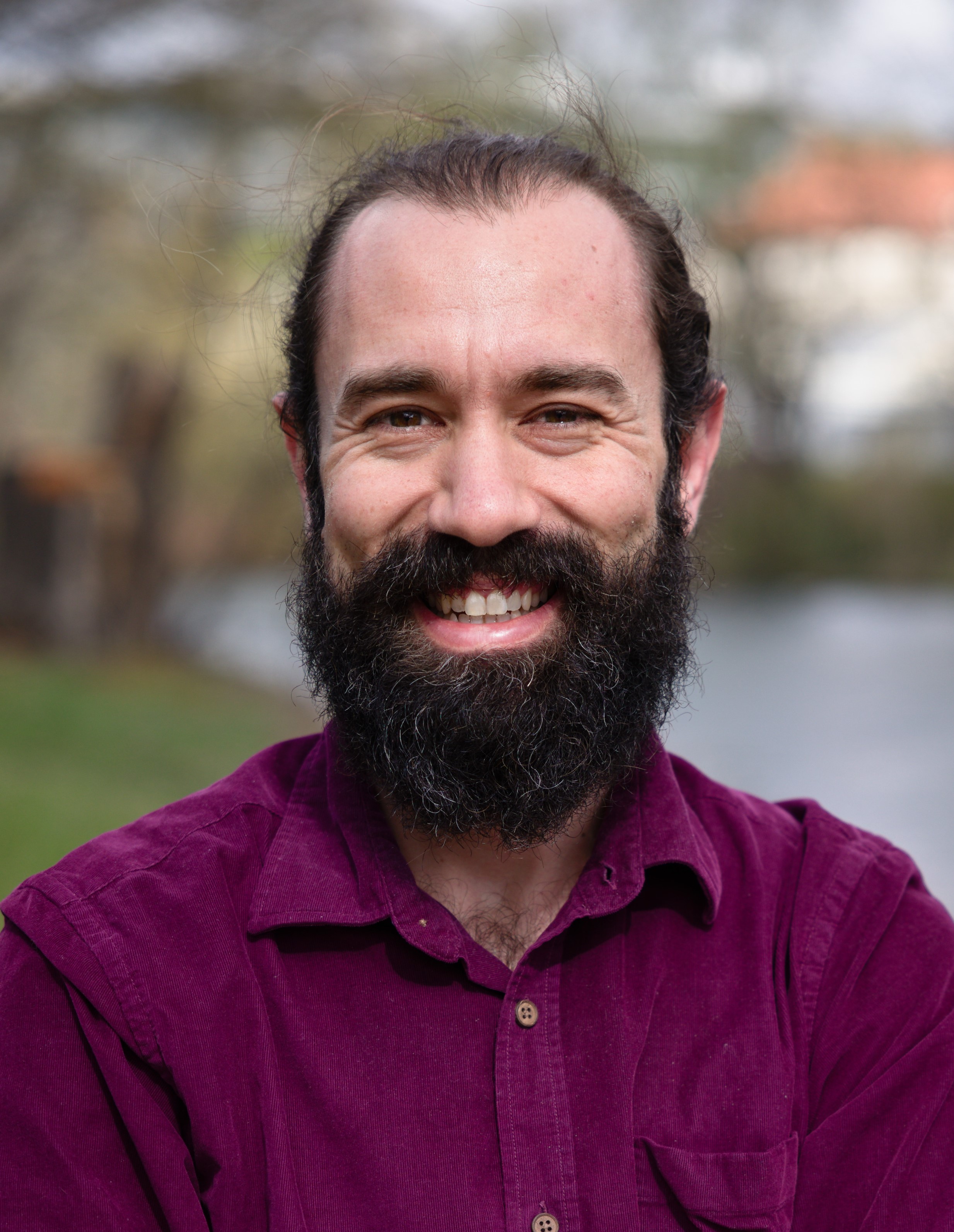}}]{Fatih Gulec} (Member, IEEE) received his B.Sc. and M.Sc. degree from Gazi University, Ankara, Turkey in 2007 and 2015, respectively both in electrical and electronics engineering. He received his Ph.D. degree in İzmir Institute of Technology, İzmir, Turkey in 2021 in electronics and communication engineering. His Ph.D. research  After a research stay as a postdoctoral researcher at the School of Electrical Engineering and Computer Science, TU Berlin, Germany with the DAAD scholarship, he is currently with the Department of Electrical Engineering and Computer Science, York University, Canada as a postdoctoral research fellow. His research interests include molecular communications and computational biology. He received the doctoral thesis award from the IEEE Turkey Section in 2023.
\end{IEEEbiography}

\begin{IEEEbiography}[{\includegraphics[width=1.1in,height=1.35in,clip,keepaspectratio]{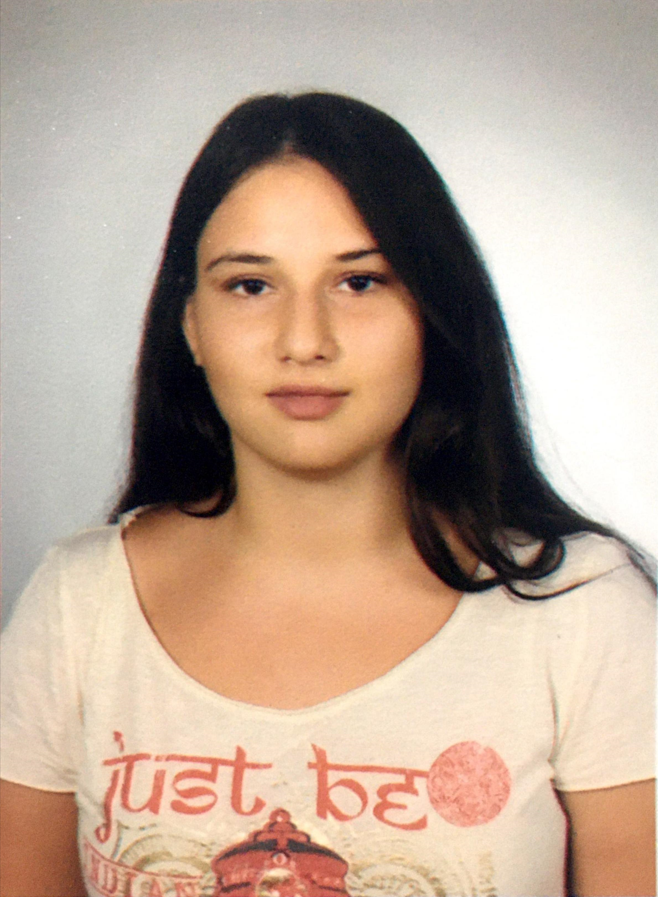}}]{Damla Yagmur Koda}  received her B.Sc. degree from Ege University, Izmir, Turkey in 2022. She is currently studying M.Sc. degree at İzmir Institute of Technology, İzmir, Turkey. She has been working as a Quality Assurance Engineer at Vestel, Manisa, Turkey since 2022. Her current research interests molecular communications and biologically inspired communications.
\end{IEEEbiography}

\begin{IEEEbiography}
[{\includegraphics[width=1.08in,height=1.4in,clip,keepaspectratio]{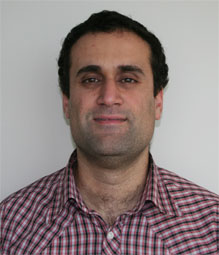}}]{Baris Atakan} received the B.Sc. degree from Ankara University, Ankara, Turkey, in 2000, the M.Sc.  degree from Middle East Technical University, Ankara, in 2005, and the Ph.D. degree from the Next-Generation and Wireless Communications Laboratory, School of Sciences and Engineering, Koç University, Istanbul, Turkey, in 2011, all in electrical and electronics engineering. He is currently a Full Professor with the Department of Electrical and Electronics Engineering, İzmir Institute of Technology, İzmir, Turkey. His current research	interests include nanoscale and molecular communications, nanonetworks and biologically inspired communications.
\end{IEEEbiography}

\begin{IEEEbiography}[{\includegraphics[width=1in,height=1.25in,clip]{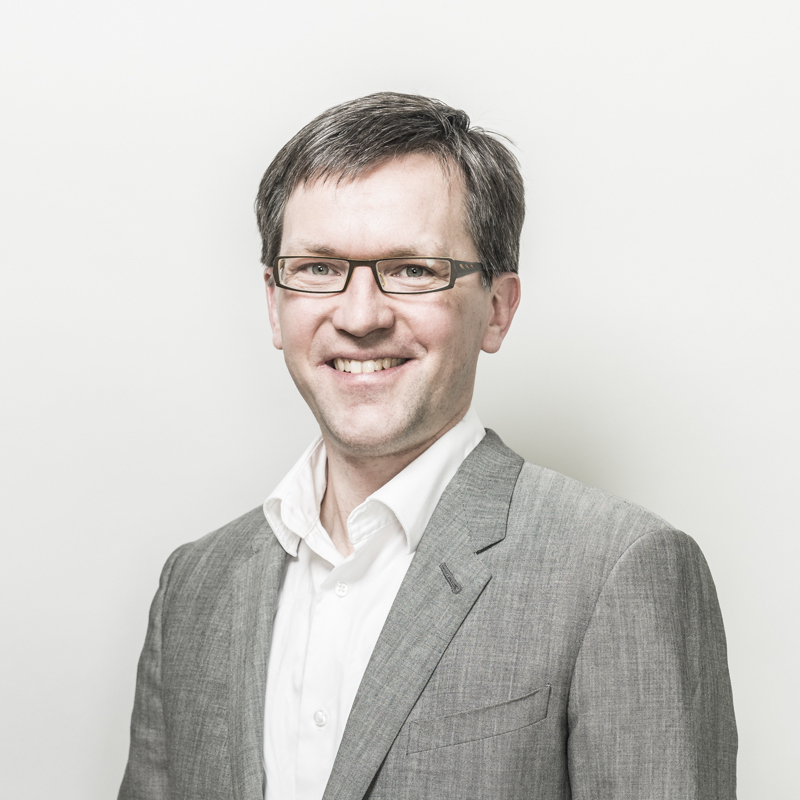}}]{Andrew W. Eckford} (Senior Member, IEEE) received the B.Eng. degree in electrical engineering from the Royal Military College of Canada in 1996, and the M.A.Sc. and Ph.D. degrees in electrical engineering from the University of Toronto in 1999 and 2004, respectively. He was a Postdoctoral Fellowship with the University of Notre Dame and the University of Toronto, prior to taking up a faculty position with York, in 2006. He is an Associate Professor with the Department of Electrical Engineering and Computer Science, York University, Toronto, ON, Canada. He has held courtesy appointments with the University of Toronto and Case Western Reserve University. In 2018, he was named a Senior Fellow of Massey College, Toronto. He is also a coauthor of the textbook Molecular Communication (Cambridge University Press). His research interests include the application of information theory to biology and the design of communication systems using molecular and biological techniques. His research has been covered in media, including The Economist, The Wall Street Journal, and IEEE Spectrum. His research received the 2015 IET Communications Innovation Award, and was a Finalist for the 2014 Bell Labs Prize.
\end{IEEEbiography}

\end{document}